\documentclass{iaus}
\usepackage{graphicx}

\def\aa{{A\&A}}

\def\aj{{AJ}}

\def\apj{{ApJ}}
\def\apjs{{ApJS}}

\def\mnras{{MNRAS}}
\def\nat{{Nature}}

\def\lsim{\mathrel{\rlap{\lower 4pt \hbox{\hskip 1pt $\sim$}}\raise 1pt
\hbox {$<$}}} 
\def\gsim{\mathrel{\rlap{\lower 4pt \hbox{\hskip 1pt $\sim$}}\raise 1pt
\hbox {$>$}}}

\newcommand{\Msun}{M_{\odot}}

\title[short title of paper] %% give here a short title %%
{Yields of Population III Supernovae and the
Abundance Patterns of Extremely Metal-Poor Stars}

\author[short author list]   %% give here a short author list %%
{K. Nomoto$^1$,
N. Tominaga$^1$,
H. Umeda$^1$,
C. Kobayashi$^2$}

\affiliation{$^1$Department of Astronomy, University of Tokyo,
  Bunkyo-ku, Tokyo 113-0033, Japan \break email:
  nomoto@astron.s.u-tokyo.ac.jp
\\[\affilskip]
$^2$Max-Planck-Institut f\"ur Astrophysik, Garching, Germany}

\pubyear{2005}
\volume{228}  %% insert here the IAU Symposium No.
\pagerange{1--7}
\date{?? and in revised form ??}
\jname{From Lithium to Uranium: Elemental Tracers of Early Cosmic Evolution}
\editors{V. Hill, P. Fran\c{c}ois \& F. Primas, eds.}
\begin{document}

\maketitle

\begin{abstract}

The abundance patterns of extremely metal-poor (EMP) stars provide us
with important information on nucleosynthesis in supernovae (SNe)
formed in a Pop III or EMP environment, and thus on the nature of the
first stars in the Universe.  We review nucleosynthesis yields of
various types of those SNe, focusing on core-collapse
(black-hole-forming) SNe with various progenitor masses, explosion
energies (including Hypernovae), and asphericity.  We discuss the
implications of the observed trends in the abundance ratios among
iron-peak elements, and the large C/Fe ratio observed in certain EMP
stars with particular attention to recently discovered hyper
metal-poor (HMP) stars.  We show that the abundance pattern of the HMP
stars with [Fe/H] $< -5$ and other EMP stars are in good accord with
those of black-hole-forming supernovae, but not pair-instability
supernovae.  This suggests that black-hole-forming supernovae made
important contributions to the early Galactic (and cosmic) chemical
evolution.  Finally we discuss the nature of First (Pop III) Stars.


\end{abstract}

\firstsection % if your document starts with a section,

\setcounter{page}{287}

\section{The First Stars and Metal-Poor Stars}

It is of vital importance to identify the first generation stars in
the Universe, i.e., totally metal-free ($Z=0$), Population (Pop) III
stars.  The impact of the formation of Pop III stars on the evolution
of the Universe depends on their typical masses.  Recent numerical
models have shown that, the first stars are as massive as $\sim$ 100
$M_\odot$ \cite{abel2002}.  The formation of long-lived low mass Pop
III stars may be inefficient because of slow cooling of metal-free gas
cloud, which is consistent with the failure of attempts to find Pop
III stars.  However, it is still controversial how the initial mass
function (IMF) depends on the metallicity.

In the early universe, the chemical enrichment by a single SN can
dominate the preexisting metal contents, because the timescale of
mixing in the interstellar medium may be longer than the age of the
Galaxy.  Thus the abundance pattern of the enriched gas may reflect
nucleosynthesis in the single SN
(\cite{aud1995,rya1996,shige1998,nak1999}).  Low-mass stars formed in
the chemically enriched gas have long lifetimes and are observed as
Extremely Metal-Poor (EMP) stars and Hyper Metal-Poor (HMP) stars
(\cite{beers2005}).  Therefore the abundance patterns of EMP and HMP
stars reflect the yield of a single SN at extremely metal-poor
environment, thus constraining the nature of the first (Pop III) SNe
and IMF.

Here we summarize the comparisons between the abundance patterns of the
EMP stars (\cite{cay2004,hon2004}) and HMP stars, HE0107--5240
(\cite{christ2002}) and HE1327--2326 (\cite{frebel2005}).

\section{Nucleosynthesis and Metallicity}

We have calculated the chemical yields for various metallicities,
including Hypernovae (HNe: see next section) (\cite{tomi2005}).

In the metal-free stellar evolution, because of the lack of initial
CNO elements, the CNO cycle dose not operate until the star contracts
to a much higher central temperature ($\sim 10^8$ K) than population
II stars, where the 3$\alpha$ reaction produces a tiny fraction of
$^{12}$C ($\sim 10^{-10}$ in mass fraction).  However, the late core
evolution and the resulting Fe core masses of metal-free stars are not
significantly different from metal-rich stars.  Therefore,
[$\alpha$/Fe] is larger by only a fraction of $\sim 0.2$ dex and the
abundance ratios of the iron-peak elements are not so different from
metal-rich stars, except for Mn.  On the other hand, the CNO cycle
produces only a small amount of $^{14}$N, which is transformed into
$^{22}$Ne during He-burning. The surplus of neutrons in $^{22}$Ne
increases the abundances of odd-Z elements (Na, Al, P, ...).
Therefore, the metallicity effect is realized for odd-Z elements and
the inverse ratio of $\alpha$ elements and their isotopes (e.g.,
$^{13}$C$/^{12}$C).  [Na/Fe] and [Al/Fe] of metal-free stars are
smaller by $\sim 1.0$ and $0.7$ dex than solar abundance stars, which
are consistent with the observed trends.

\begin{figure*}[t]
\centering
\resizebox{120mm}{!}{\includegraphics{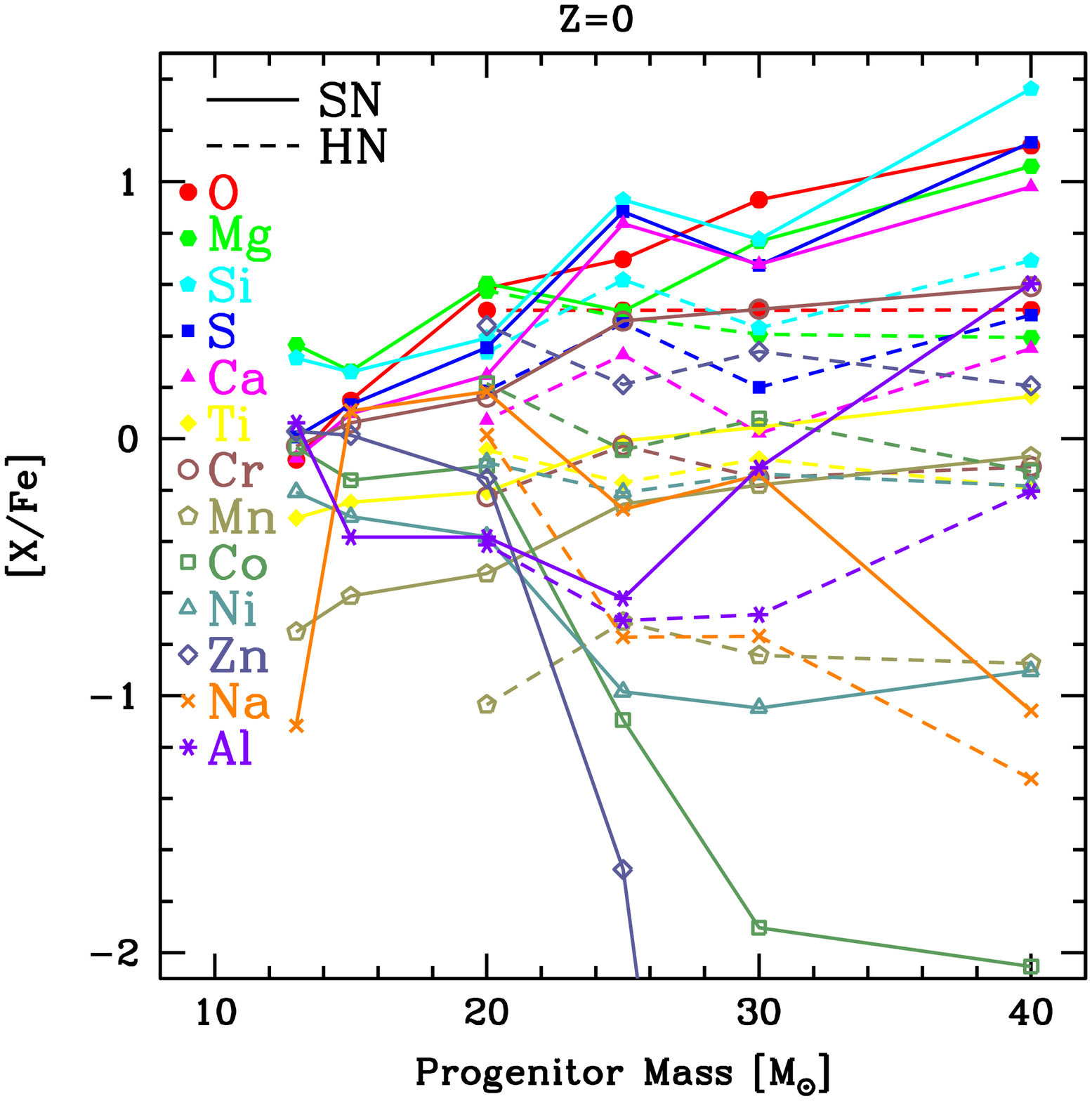}\includegraphics{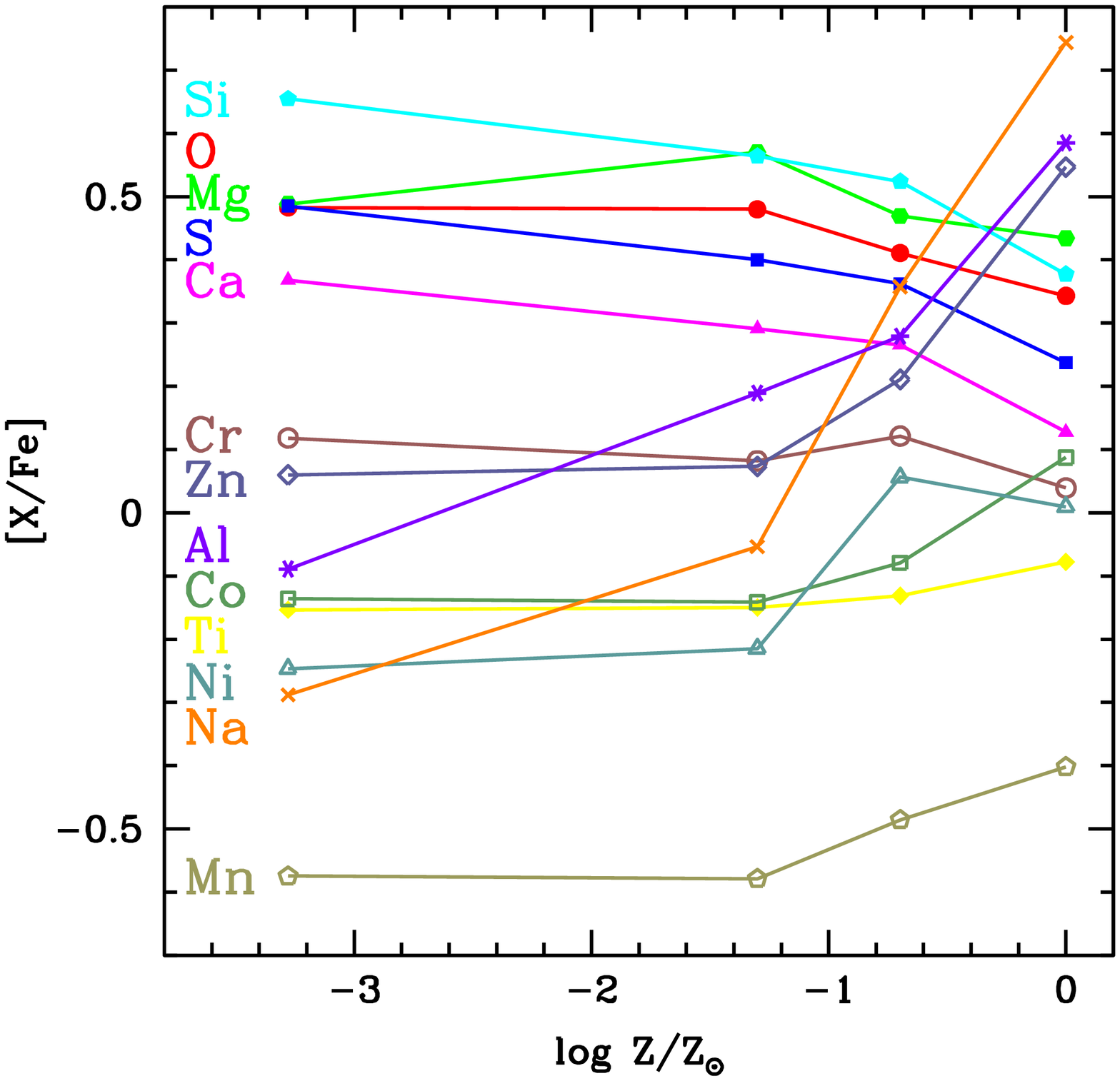}}
\caption{(left) Relative abundance ratios as a function of progenitor 
mass with $Z=0$.  The solid and dashed lines show normal SNe II with
$E_{51}=1$ and HNe.  (right) The IMF weighted abundance ratios as a
function of metallicity of progenitors, where the HN fraction
$\epsilon_{\rm HN}=0.5$ is adopted.  $Z=0$ results are plotted at
$\log Z/Z_\odot=-4$.}
\label{fig:yield-z}
\end{figure*}

Figure \ref{fig:yield-z} shows the abundance ratios of Pop III ($Z=0$)
SNe II and HNe as a function of the progenitor mass (left) and the IMF
weighted yields of SNe II and HNe as a function of metallicity
(right).

The solid and dashed lines show the SN II and HN yields, respectively.
The yield masses of $\alpha$ elements (O, Ne, Mg, Si, S, Ar, Ca, and
Ti) are larger for more massive stars because of the larger mantle
mass.  Since the Fe mass is $\sim 0.1M_\odot$, being independent of
the progenitor's mass for $E_{51}$=1, the abundance ratio
[$\alpha$/Fe] is larger for more massive stars.  For HNe, although Fe
production is larger for more massive stars because of the higher
energy, [$\alpha$/Fe] is almost constant independent of the stellar
mass because we assume the mass-cut to get [O/Fe] $=0.5$.

\section{Hypernova Yields}

Regarding supernova (SN) yields, one of the most interesting recent
developments is the discovery of hyper-energetic supernovae, whose
kinetic energy (KE) exceeds $10^{52}$\,erg, about 10 times the KE of
normal core-collapse SNe (hereafter $E_{51} =
E/10^{51}$\,erg) (\cite{nomoto2004}).  Such SNe are called Hypernovae
(HNe).  Nucleosynthesis in HNe has some special features, so that the
contribution of HNe in early chemical evolution can be seen the
abundance patterns of HMP/EMP stars.

\begin{figure*}[t]
\resizebox{\hsize}{!}{\includegraphics{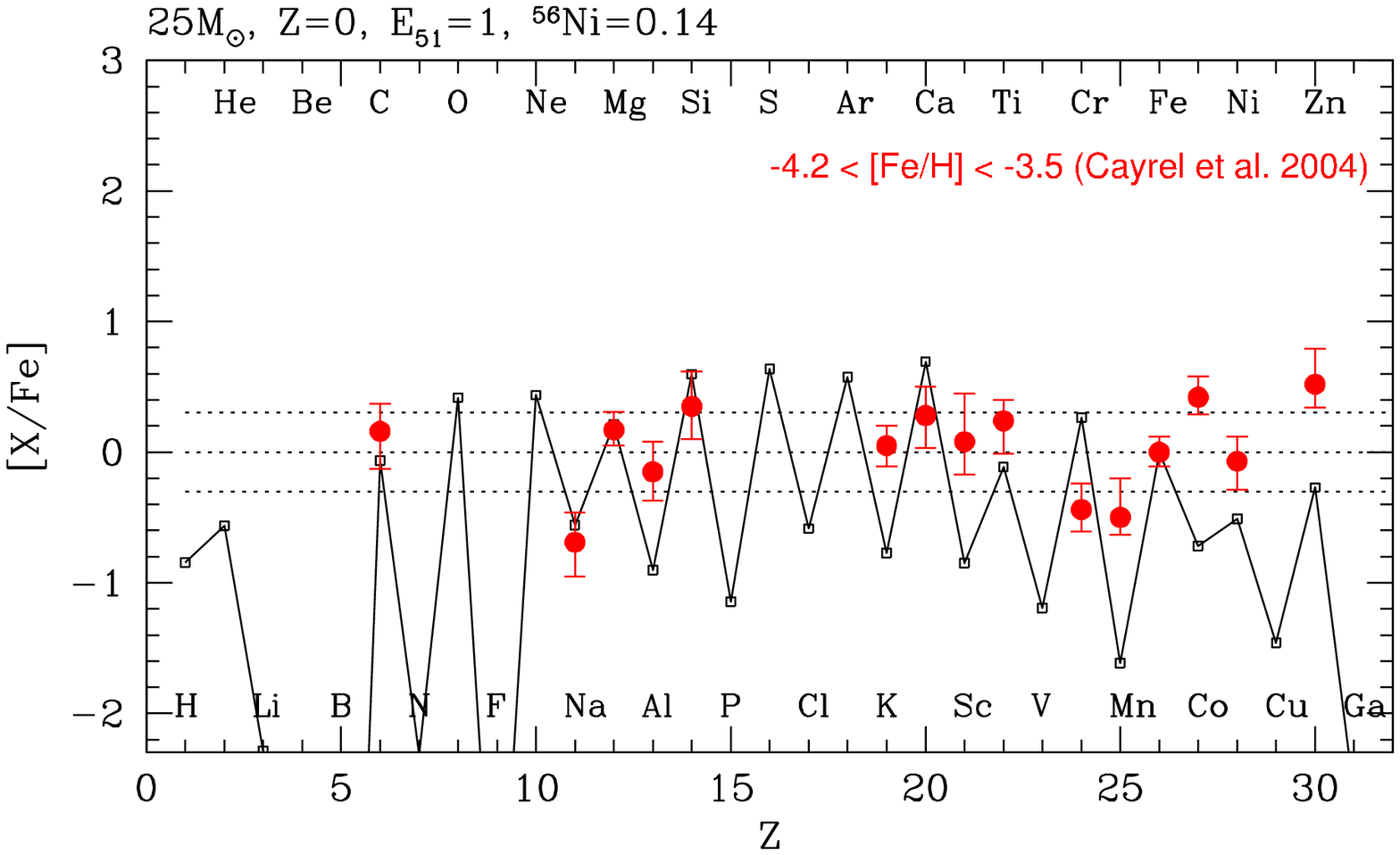}
\includegraphics{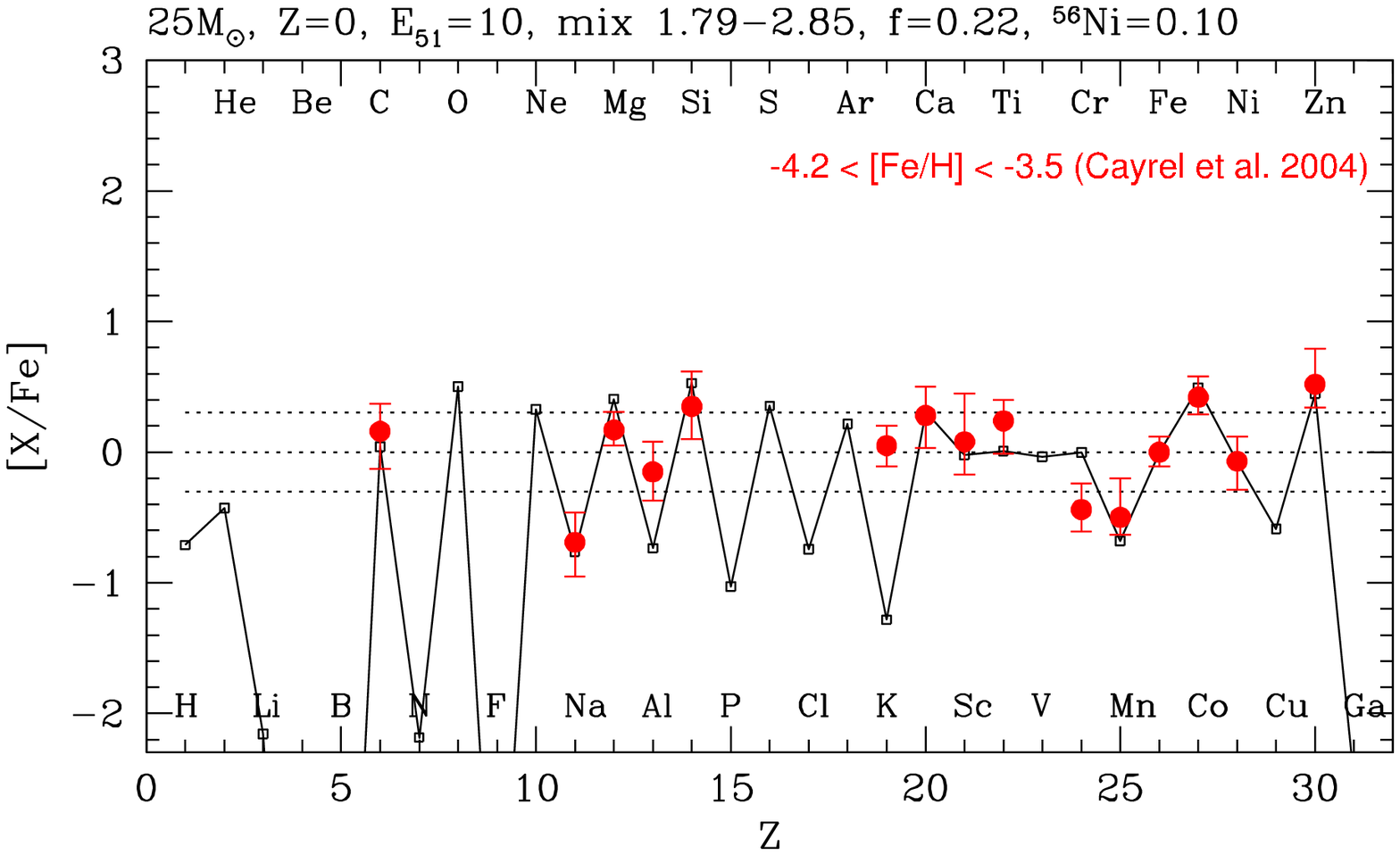}}
\caption{Averaged elemental abundances of stars with [Fe/H] $= -3.7$
(Cayrel et al. 2004) compared with the normal SN yield (left: 25
$M_\odot$, $E_{51} =$ 1), and the hypernova yield (right: 25
$M_\odot$, $E_{51} =$ 10).}
\label{fig7}
\end{figure*}

\subsection{Hypernovae and Faint Supernovae}

The Hypernovae 1998bw and 2003dh were clearly linked to the
Gamma-Ray Bursts GRB 980425 (Galama et al. 1998) and GRB 030329
(Stanek et al. 2003; Hjorth et al. 2003), thus establishing the
connection between long GRBs and core-collapse supernovae (SNe).
SNe~1998bw and 2003dh were exceptional for SNe~Ic: they were as
luminous at peak as a SN~Ia, indicating that they synthesized 0.3 -
0.5 $M_\odot$ of $^{56}$Ni, and their kinetic energy (KE) were
estimated as $E_{51} \sim$ 30 - 50 (Iwamoto et al. 1998; Woosley et
al. 1999; Nakamura et al. 2001a).

Other HNe have been recognized, such as SN~1997ef (Iwamoto et
al. 2000; Mazzali, Iwamoto, \& Nomoto 2000) and SN~2002ap (Mazzali et
al. 2002).  These hypernovae span a wide range of properties, although
they all appear to be highly energetic compared to normal
core-collapse SNe.  The mass estimates, obtained from fitting the
optical light curves and spectra, place hypernovae at the high-mass
end of SN progenitors.

In contrast, SNe II 1997D and 1999br were very faint SNe with very low
KE (Turatto et al. 1998; Hamuy 2003; Zampieri et al. 2003).  In the
diagram that shows $E$ and the mass of $^{56}$Ni ejected $M(^{56}$Ni)
as a function of the main-sequence mass $M_{\rm ms}$ of the progenitor
star, therefore, we propose that SNe from stars with $M_{\rm ms}
\mathrel{\rlap{\lower 4pt \hbox{\hskip 1pt $\sim$}}\raise 1pt \hbox
{$>$}}$ 20-25 $M_\odot$ have different $E$ and $M(^{56}$Ni), with a
bright, energetic ``hypernova branch'' at one extreme and a faint,
low-energy SN branch at the other (Nomoto et al. 2003).  For the faint
SNe, the explosion energy was so small that most $^{56}$Ni fell back
onto the compact remnant.  The extreme cases of the faint SN branch
may correspond to HMP stars as will be discussed below.  Between the
two branches, there may be a variety of SNe (Hamuy 2003).

This trend might be interpreted as follows.  Stars more massive than
$\sim$ 25 $M_\odot$ form a black hole at the end of their evolution.
Stars with non-rotating black holes are likely to collapse "quietly"
ejecting a small amount of heavy elements (Faint supernovae).  In
contrast, stars with rotating black holes are likely to give rise to
Hypernovae.  The hypernova progenitors might form the rapidly rotating
cores by spiraling-in of a companion star in a binary system.

\subsection{Nucleosynthesis in Hypernovae}

In core-collapse supernovae/hypernovae, stellar material undergoes
shock heating and subsequent explosive nucleosynthesis. Iron-peak
elements are produced in two distinct regions, which are characterized
by the peak temperature, $T_{\rm peak}$, of the shocked material.  For
$T_{\rm peak} > 5\times 10^9$K, material undergoes complete Si burning
whose products include Co, Zn, V, and some Cr after radioactive
decays.  For $4\times 10^9$K $<T_{\rm peak} < 5\times 10^9$K,
incomplete Si burning takes place and its after decay products include
Cr and Mn (e.g., Nakamura et al. 1999).

We note the following characteristics of nucleosynthesis with very
large explosion energies (Nakamura et al. 2001b; Nomoto et al. 2001;
Umeda \& Nomoto 2005):

(1) Both complete and incomplete Si-burning regions shift outward in
mass compared with normal supernovae, so that the mass ratio between
the complete and incomplete Si-burning regions becomes larger.  As a
result, higher energy explosions tend to produce larger [(Zn, Co,
V)/Fe] and smaller [(Mn, Cr)/Fe], which can explain the trend observed
in very metal-poor stars (Umeda \& Nomoto 2005).

(2) In the complete Si-burning region of hypernovae, elements produced
by $\alpha$-rich freezeout are enhanced.  Hence, elements synthesized
through capturing of $\alpha$-particles, such as $^{44}$Ti, $^{48}$Cr,
and $^{64}$Ge (decaying into $^{44}$Ca, $^{48}$Ti, and $^{64}$Zn,
respectively) are more abundant.

Hypernova nucleosynthesis may have made an important contribution to
Galactic chemical evolution.  In the early galactic epoch when the
galaxy was not yet chemically well-mixed, [Fe/H] may well be
determined by mostly a single SN event (Audouze \& Silk 1995). The
formation of metal-poor stars is supposed to be driven by a supernova
shock, so that [Fe/H] is determined by the ejected Fe mass and the
amount of circumstellar hydrogen swept-up by the shock wave (Ryan,
Norris, \& Beers 1996).  Then, hypernovae with larger $E$ are likely
to induce the formation of stars with smaller [Fe/H], because the mass
of interstellar hydrogen swept up by a hypernova is roughly
proportional to $E$ (Ryan et al. 1996; Shigeyama \& Tsujimoto 1998)
and the ratio of the ejected iron mass to $E$ is smaller for
hypernovae than for normal supernovae.

\begin{figure*}[t]
\centering
\resizebox{105mm}{!}{\includegraphics{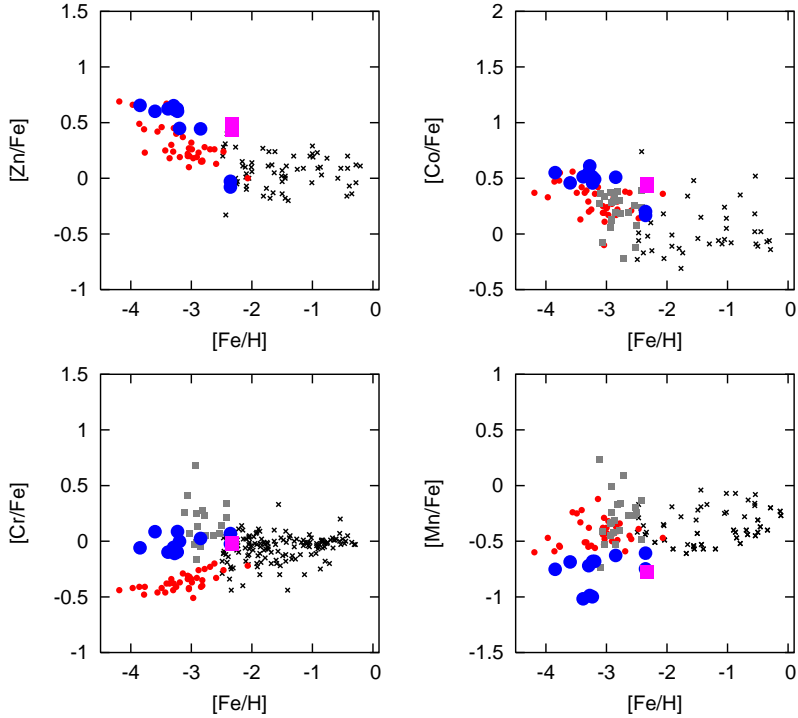}}
\caption{The comparison between the [Zn, Co, Cr, Mn/Fe] trends of
observed stars (the previous studies {\it cross}, Cayrel et al. (2004)
{\it red circle}, Honda et al. (2004) {\it gray square}), and those of
individual stars models ({\it blue circle}) and IMF integration ({\it
pink square}).}
\label{fig4c}
\end{figure*}

\section{Extremely Metal-Poor (EMP) Stars}

Here we make comparisons between the theoretical yields and the
abundance patterns of the EMP stars, given in Cayrel et al. (2004) and
Honda et al. (2004) where the data of 35 and 22 metal-poor stars are
provided, respectively.  Though the trends of elemental abundance
ratios are similar to those found earlier (\cite{mcw1995,rya1996}),
Cayrel et al. (2004) found much smaller dispersions.

\subsection{Averaged Abundance Patterns of EMP Stars}

First the theoretical yields are compared with the averaged abundance
pattern of four EMP stars, CS~22189-009, CD-38:245, CS~22172-002 and
CS~22885-096, which have low metallicities ($-4.2<{\rm [Fe/H]}<-3.5$)
and normal [C/Fe] $\sim 0$ (\cite{cay2004}).  Figures~\ref{fig7} show
the comparison with the 25 $M_\odot$ model of normal energy explosion
($E_{51} = 1$: left) and hyper energetic explosion ($E_{51} = 10$:
right).

In the normal SN model, the mass-cut is determined to eject Fe of mass
0.14 $M_\odot$).  Then the yields are in reasonable agreements with
the observations for the ratios [(Na, Mg, Si)/Fe], but give too small
[(Sc, Ti, Mn, Co, Ni, Zn)/Fe] and too large [(Ca, Cr)/Fe].

In the HN model (right), these ratios are in much better agreement
with observations.  The ratios of Co/Fe and Zn/Fe are larger in higher
energy explosions since both Co and Zn are synthesized in complete Si
burning at high temperature region.  To account for the observations,
materials synthesized in a deeper complete Si-burning region should be
ejected, but the amount of Fe should be small.  This is realized in
the mixing-fallback models (Umeda \& Nomoto 2002, 2005).

In the HN model (right), the low-density progenitor model is used
(\cite{umeda2005}) to enhance the $\alpha$-rich freeze-out.  As a
result, the Sc/Fe and Ti/Fe ratios are especially enhanced.

\subsection{Hypernovae and the Abundance Trends}

It has been found that, in the observed abundances of halo stars,
there are significant differences between the abundance patterns in
the iron-peak elements below and above [Fe/H]$ \sim -2.5$ - $-3$.

(1) For [Fe/H]$\lsim -2.5$, the mean values of [Cr/Fe] and [Mn/Fe]
decrease toward smaller metallicity, while [Co/Fe] increases
(McWilliam et al. 1995; Ryan et al. 1996).  Cayrel et al. (2004) found
much flatter trend of [Mn/Fe].

(2) [Zn/Fe]$ \sim 0$ for [Fe/H] $\simeq -3$ to $0$ (Sneden, Gratton,
\& Crocker 1991), while at [Fe/H] $< -3.3$, [Zn/Fe] increases toward
smaller metallicity (Cayrel et al. 2004).

The larger [(Zn, Co)/Fe] and smaller [(Mn, Cr)/Fe] in the supernova
ejecta can be realized if the mass ratio between the complete Si
burning region and the incomplete Si burning region is larger, or
equivalently if deep material from the complete Si-burning region is
ejected by mixing or aspherical effects.  This can be realized if (1)
the mass cut between the ejecta and the compact remnant is located at
smaller $M_r$ (Nakamura et al. 1999), (2) $E$ is larger to move the
outer edge of the complete Si burning region to larger $M_r$ (Nakamura
et al. 2001b), or (3) asphericity in the explosion is larger.

Among these possibilities, a large explosion energy $E$ enhances
$\alpha$-rich freezeout, which results in an increase of the local
mass fractions of Zn and Co, while Cr and Mn are not enhanced (Umeda
\& Nomoto 2002).  Models with $E_{51} = 1 $ do not produce
sufficiently large [Zn/Fe].  To be compatible with the observations of
[Zn/Fe] $\sim 0.5$, the explosion energy must be much larger, i.e.,
$E_{51} \gsim 20$ for $M \gsim 20 M_\odot$, i.e., hypernova-like
explosions of massive stars ($M \gsim 25 M_\odot$) with $E_{51} > 10$
are responsible for the production of Zn.  

In the hypernova models, moreover, the overproduction of Ni, as found
in the simple ``deep'' mass-cut model (1), can be avoided (Umeda \&
Nomoto 2005).  Therefore, if hypernovae made significant contributions
to the early Galactic chemical evolution, it could explain the large
Zn and Co abundances and the small Mn and Cr abundances observed in
very metal-poor stars.

\begin{figure*}[t]
\centering
\resizebox{105mm}{!}{\includegraphics{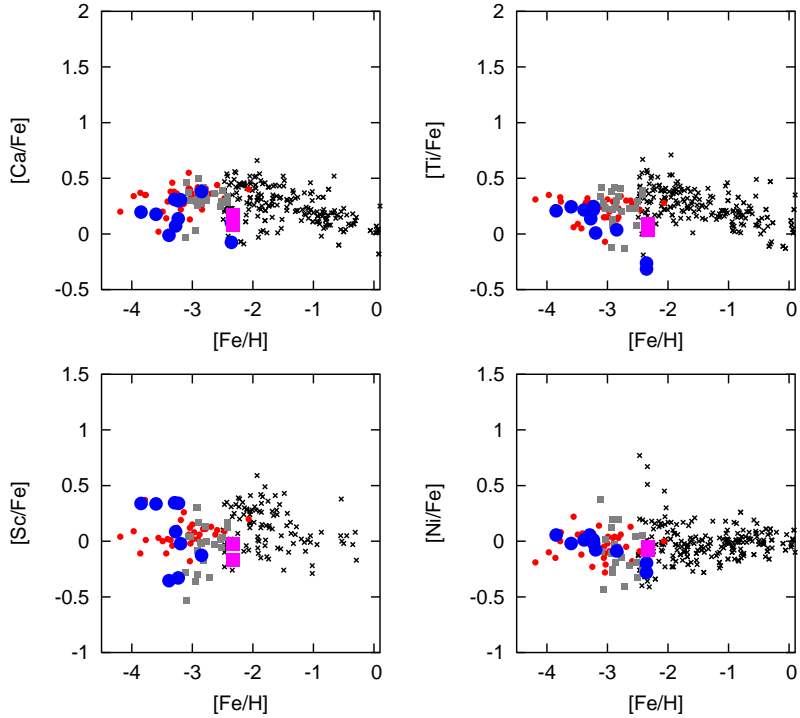}}
\caption{The comparison between the [Ca, Ti, Sc, Ni/Fe] trends of
observed stars (the previous studies {\it cross}, Cayrel et al. (2004)
{\it red circle}, Honda et al. (2004) {\it gray square}) and those of
individual stars models ({\it blue circle}) and IMF integration ({\it
pink square}).}
\label{fig4b}
\end{figure*}

\begin{figure*}[t]
\centering
\resizebox{105mm}{!}{\includegraphics{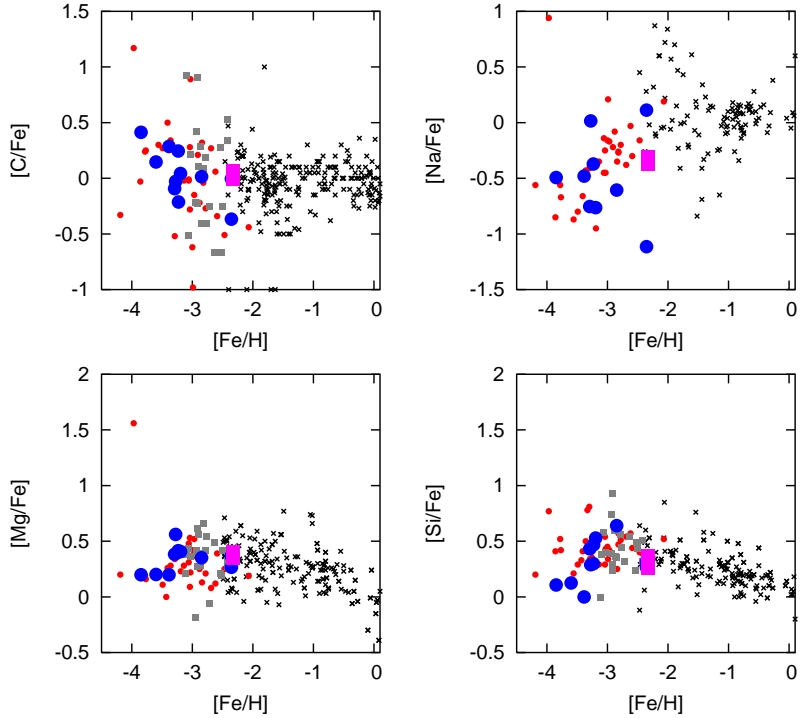}}
\caption{The comparison between the [C, Na, Mg, Si/Fe] trends of
observed stars (the previous studies {\it cross}, Cayrel et al. (2004)
{\it red circle}, and Honda et al. (2004) {\it gray square}) and those
of individual stars models ({\it blue circle}) and IMF integration
({\it pink square}).}
\label{fig4a}
\end{figure*}

 Figures \ref{fig4c}-\ref{fig4a} show the theoretical trial to
reproduce the observed trend in Cayrel et al. (2004) and Honda et
al. (2004) with the supernova-induced star formation model.  The
supernova models with $(M/M_\odot, E_{51}) =$ (13, 1), (15, 1),
(20, 10), (25, 10), (30, 20), (40, 30), and (50, 40) are used,
where the stars more massive than 20 $\Msun$ are assumed to explode as
HNe.  These models are located at [Fe/H] = log$_{10}$ (Fe/$E_{51})-C$,
and can be consistent with the observed trend even without mixing of
materials in the interstellar medium.  Since HNe may explode as
jet-like explosions, the HN models adopt the mixing-fallback and the
``low-density'' model (see Umeda \& Nomoto 2005; Tominaga et al. 2005
for details).

The trends of most elements can be well reproduced by these models,
except for the overproduction of Cr and the underproduction of K.  For
Cr, only the gradient of Cr with [Fe/H] can be reproduced.

\section{Hyper Metal-Poor (HMP) Stars}

Recently two HMP stars, HE0107--5240 (\cite{christ2002}) and
HE1327--2326 (\cite{frebel2005}), were discovered, whose metallicity
Fe/H is smaller than 1/100,000 of the Sun (i.e., [Fe/H] $< -5$), being
more than a factor of 10 smaller than previously known EMP stars.
These discoveries have raised an important question as to whether the
observed low mass ($\sim$ 0.8~$M_\odot$) HMP stars are actually Pop
III stars, or whether these HMP stars are the second generation stars
being formed from gases which were chemically enriched by a single
first generation supernova (SN) (\cite{umeda2003}).  This is related
to the questions of how the initial mass function depends on the
metallicity.  Thus identifying the origin of these HMP stars is
indispensable to the understanding of the earliest star formation and
chemical enrichment history of the Universe.

\subsection{Abundance Patterns of HE0107--5240 \& HE1327--2326}

\begin{figure*}[t]
\centering
\resizebox{130mm}{!}{\includegraphics{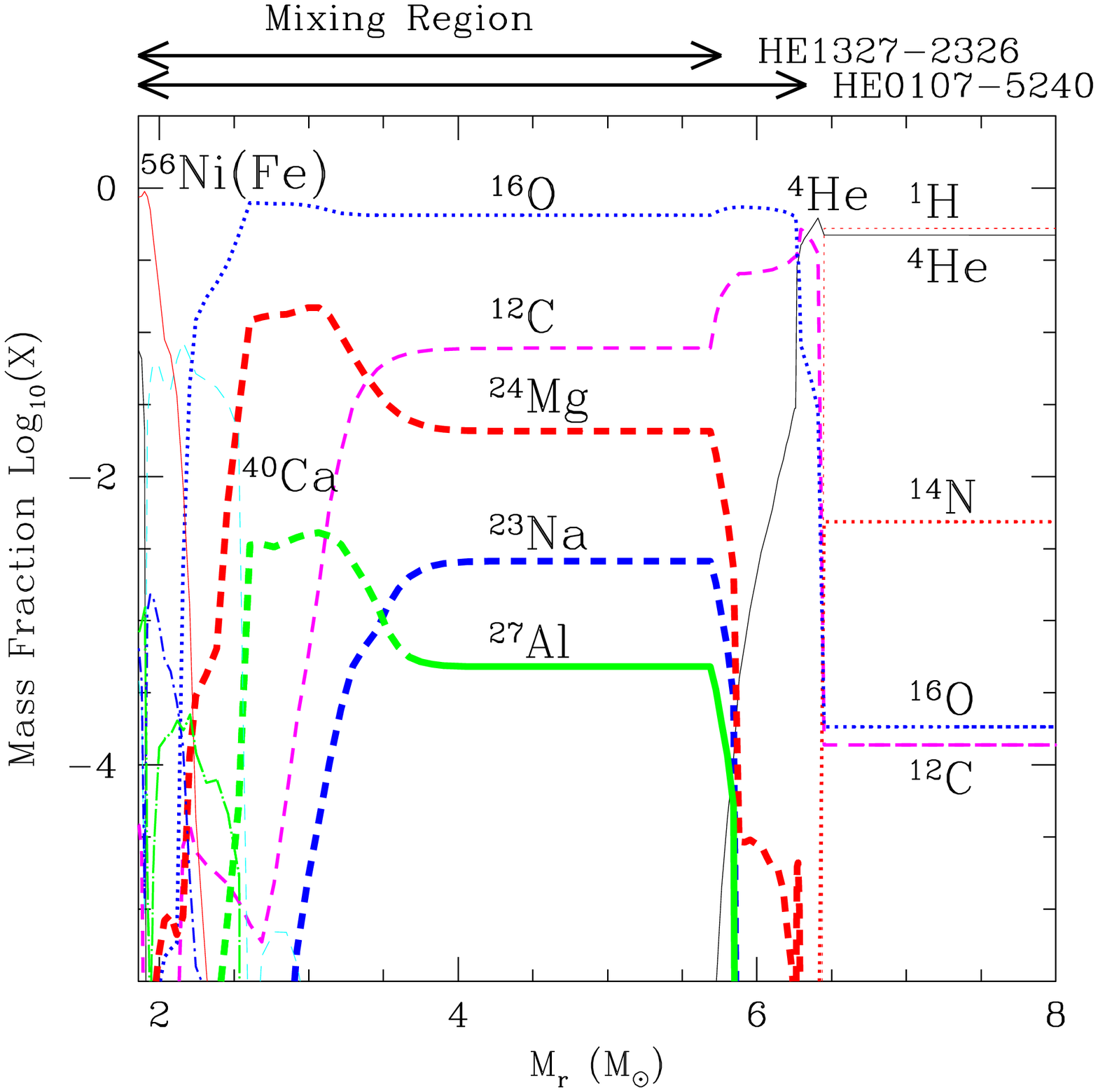}
\includegraphics{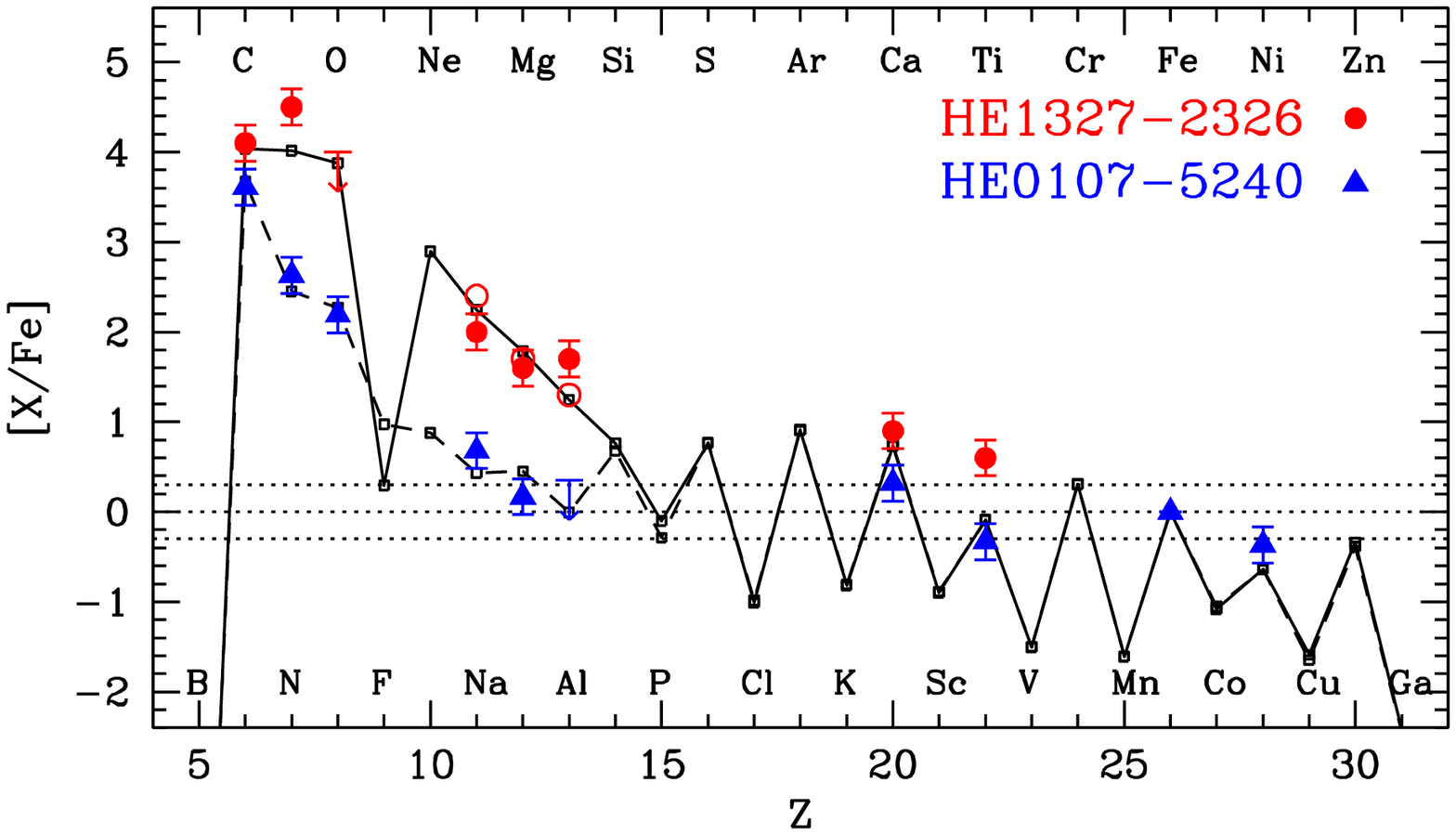}
}
\caption{
(left) The post-explosion abundance distributions for the 25 $M_\odot$
model with the explosion energy $E_{51} =$ 0.7 (Iwamoto et al. 2005).
(right) Elemental abundances of HMP stars (filled circles and
triangle, compared with a theoretical supernova yield (Iwamoto et al.
2003, 2005).}
\label{fig5}
\end{figure*}

 The elemental abundance patterns of these HMP stars provide a key to
the answer to the above questions.  The abundance patterns of
HE1327--2326 and HE0107--5240 are quite unusual (Fig.~\ref{fig5}). The
striking similarity of [Fe/H] (=$-5.4$ and $-5.2$ for HE1327--2326 and
HE0107--5240, respectively) and [C/Fe] ($\sim +4$) suggests that
similar chemical enrichment mechanisms operated in forming these HMP
stars.  However, the N/C and (Na, Mg, Al)/Fe ratios are more than a
factor of 10 larger in HE1327--2326.  In order for the theoretical
models to be viable, these similarities and differences should be
explained self-consistently.

Iwamoto et al. (2005) showed that the above similarities and
variations of the HMP stars can be well reproduced in unified manner
by nucleosynthesis in the core-collapse ``faint'' supernovae (SNe)
which undergo mixing-and-fallback (\cite{umeda2003}).  We thus argue
that the HMP stars are the second generation low mass stars, whose
formation was induced by the first generation (Pop III) SN with
efficient cooling of carbon-enriched gases.

\subsection{Models for HE0107--5240 \& HE1327--2326}

We consider a model that C-rich EMP stars are produced in the ejecta
of (almost) metal-free supernova mixed with extremely metal-poor
interstellar matter.  The similarity of [Fe/H] and [C/Fe] suggests
that the progenitor's masses of Pop III SNe were similar for these HMP
stars.  We therefore choose the Pop III 25 $M_\odot$ models and
calculate their evolution and explosion 
(\cite{umeda2003,iwamoto2005}).

The abundance distribution after explosive nucleosynthesis is shown in
Figure~\ref{fig5} (left) for the kinetic energy $E$ of the ejecta
$E_{51} \equiv E/10^{51}~{\rm erg} = 0.74$.  The abundance
distribution for $E_{51} = 0.71$ is similar.  In the ``faint'' SN
model, most part of materials that underwent explosive nucleosynthesis
are decelerated by the influence of the gravitational pull
(\cite{woosley1995}) and will eventually fall back onto the central
compact object.  The explosion energies of $E_{51} = 0.74$ and $0.71$
lead to the mass cut $M_{\rm cut} = 5.8 M_\odot$ and $6.3 M_\odot$,
respectively, and the former and the latter models are used to explain
the abundance patterns of HE1327--2326 and HE0107--5240, respectively.

 During the explosion, the SN ejecta is assumed to undergo mixing,
i.e., materials are first uniformly mixed in the mixing-region
extending from $M_r = 1.9 M_\odot$ to the mass cut at $M_r = M_{\rm
cut}$ (where $M_r$ is the mass coordinate and stands for the mass
interior to the radius $r$) as indicated in Figure~\ref{fig5} (left),
and only a tiny fraction, $f$, of the mixed material is ejected from
the mixing-region together with all materials at $M_r > M_{\rm cut}$;
most materials interior to the mass cut fall back onto the central
compact object.  Such a mixing-fallback mechanism (which might mimic a
jet-like explosion) is required to extract Fe-peak and other heavy
elements from the deep fallback region into the
ejecta (\cite{umeda2003,umeda2005}).

 Figure~\ref{fig5} (right) shows the calculated abundance ratios in
the SN ejecta models for suitable choice of $f$ which are respectively
compared with the observed abundances of the two HMP stars.  To
reproduce [C/Fe] $\sim$ +4 and other abundance ratios of HMP stars in
Figure~\ref{fig5} (right), the ejected mass of Fe is only 1.0 $\times
10^{-5}M_\odot$ for HE1327--2326 and 1.4 $\times 10^{-5}M_\odot$ for
HE0107--5240.  These SNe are much fainter in the radioactive tail than
the typical SNe and form massive black holes of $\sim 6 M_\odot$.

 The question is what causes the large difference in the amount of
Na-Mg-Al between the SNe that produced HE0107--5240 and HE1327--2326.
Because very little Na-Mg-Al is ejected from the mixed
fallback materials (i.e., $f \sim 10^{-4}$) compared with the
materials exterior to the mass cut, the ejected amount of Na-Mg-Al is
very sensitive to the location of the mass cut.  As indicated in
Figure~6, $M_{\rm cut}$ is smaller (i.e., the fallback mass is
smaller) in the model for HE1327--2326 ($M_{\rm cut} = 5.8 M_\odot$)
than HE0107--5240 ($M_{\rm cut} = 6.3 M_\odot$), so that a larger
amount of Na-Mg-Al is ejected from the SN for HE1327--2326.  Since
$M_{\rm cut}$ is sensitively determined by the explosion energy, the
(Na-Mg-Al)/Fe ratios among the HMP stars are predicted to show
significant variations and can be used to constrain $E_{51}$.  
Note also that the explosion energies of these SN models with fallback
are not necessarily very small (i.e., $E_{51} \sim 0.7$).
Further these explosion energies are consistent with those observed in
the actual ``faint'' SNe (\cite{turatto1998}).

 The next question is why HE1327--2326 has a much larger N/C ratio
than HE0107--5240.  In our models, a significant amount of N is
produced by the mixing between the He convective shell and the H-rich
envelope during the presupernova evolution (\cite{umeda2000}), where C
created by the triple-$\alpha$ reaction is burnt into N through the CN
cycle.  For the HE1327--2326 model, we assume about 30 times larger
diffusion coefficients (i.e., faster mixing) for the H and He
convective shells to overcome an inhibiting effect of the mean
molecular weight gradient (and also entropy gradient) between H and He
layers.  Thus, larger amounts of protons are carried into the He
convective shell.  Then [C/N] $\sim 0$ is realized as observed in
HE1327--2326.  Such an enhancement of mixing efficiency has been
suggested to take place in the present-day massive stars known as fast
rotators, which show various N and He enrichments due to different
rotation velocities (\cite{heger2002}).

\section{First Stars}

Recent numerical models have shown that, the first stars are as
massive as $\sim$ 100 $M_\odot$ (Abel et al. 2002).  The formation of
long-lived low mass Pop III stars may be inefficient because of slow
cooling of metal free gas cloud, which is consistent with the failure
of attempts to find Pop III stars.

If the most Fe deficient star, HE0107-5240, is a Pop III low mass star
that has gained its metal from a companion star or interstellar matter
(Yoshii 1981), would it mean that the above theoretical arguments
are incorrect and that such low mass Pop III stars have not been
discovered only because of the difficulty in the observations?

Based on the results in the earlier section, we suggest that the first
generation supernovae were the explosion of $\sim$ 20-130 $M_\odot$
stars and some of them produced C-rich, Fe-poor ejecta.  Then the low
mass star with even [Fe/H] $< -5$ can form from the gases of mixture
of such a supernova ejecta and the (almost) metal-free interstellar
matter, because the gases can be efficiently cooled by enhanced C and
O ([C/H] $\sim -1$).

In contrast to the core-collapse supernovae of 20-130 $M_\odot$ stars,
the observed abundance patterns cannot be explained by the explosions
of more massive, 130 - 300 $M_\odot$ stars. These stars undergo
pair-instability supernovae (PISNe) and are disrupted completely
(e.g., Umeda \& Nomoto 2002; Heger \& Woosley 2002), which cannot be
consistent with the large C/Fe observed in HMP stars and other
C-rich EMP stars.  The abundance ratios of iron-peak elements ([Zn/Fe]
$< -0.8$ and [Co/Fe] $< -0.2$) in the PISN ejecta (Fig.~\ref{fig7};
Umeda \& Nomoto 2002; Heger \& Woosley 2002) cannot explain the large
Zn/Fe and Co/Fe in the typical EMP stars (McWilliam et al. 1995;
Norris et al. 2001; Cayrel et al. 2003) and CS22949-037 either.
Therefore the supernova progenitors that are responsible for the
formation of EMP stars are most likely in the range of $M \sim 20 -
130$ $M_\odot$, but not more massive than 130 $M_\odot$.  This upper
limit depends on the stability of massive stars.

%\appendix

%\begin{acknowledgments}
%We would like to acknowledge the useful comments of a referee concerning
%the solution procedure used in \S\,\ref{sec:concl}. A.\,N.\,O. is supported
%by SERC under grant number GR/F/12345.
%\end{acknowledgments}

\end{document}